\documentclass[aps,twocolumn,pra,
preprintnumbers,amsmath,amssymb,superscriptaddress]{revtex4}
\usepackage{graphicx}
\usepackage{color}

\definecolor{refkey}{rgb}{0.9451,0.2706,0.4941}
\definecolor{labelkey}{rgb}{0.9451,0.2706,0.4941}

\newcommand{\bea}{\begin{align}}
\newcommand{\eea}{\end{align}}

\newcommand{\mc}[1]{\mathcal{#1}}

\newcommand{\Lbar}[1]{\overline{#1}}

\begin{document}

\preprint{OU-HET-864}
\preprint{RIKEN-STAMP-14}
\preprint{WIS/06/15-Oct-DPPA}
\preprint{YITP-15-64}

\title{Exact Path Integral for 3D Quantum Gravity II}

\date{\today}

\author{Masazumi Honda}\email[]{masazumi.honda@weizmann.ac.il} 
\affiliation{{\it Department of Particle Physics, Weizmann Institute of Science, Rehovot 7610001, Israel}}

\author{Norihiro Iizuka}\email[]{iizuka@phys.sci.osaka-u.ac.jp} 
\affiliation{{\it Department of Physics, Osaka University, Toyonaka, Osaka 560-0043, JAPAN}}

\author{Akinori Tanaka}
\email[]{akinori.tanaka@riken.jp}
\affiliation{{\it Interdisciplinary Theoretical Science Research Group, 
RIKEN, Wako 351-0198, JAPAN}}

\author{Seiji Terashima}
\email[]{terasima@yukawa.kyoto-u.ac.jp}
\affiliation{{\it Yukawa Institute for Theoretical Physics, 
Kyoto University, Kyoto 606-8502, JAPAN}}

\begin{abstract}
Continuing the work arXiv:1504.05991, we discuss various aspects of three dimensional quantum gravity partition function in AdS in the semi-classical limit. The partition function is holomorphic and is the one which we obtained by using the localization technique of Chern-Simons theory in arXiv:1504.05991. We obtain a good expression for it in the summation form over Virasoro characters for the vacuum and primaries. A key ingredient for that is an interpretation of boundary localized fermion. We also check that the coefficients in the summation form over Virasoro characters of the partition function are positive integers and satisfy the Cardy formula. These give physical interpretation that these coefficients represent the number of primary fields in the dual CFT in the large $k$ limit. 

\end{abstract}

\maketitle



\noindent
\section{Introduction} 
Solving and writing down an explicit form of the partition function for quantum gravity is one of the most important remaining problems in 
theoretical physics.  
In \cite{Iizuka:2015jma}, 
the authors write down the explicit partition function for 3D quantum pure 
gravity in asymptotic AdS spacetime. 
The partition function is the direct product of holomorphic and anti-holomorphic function, 
and is obtained by using the Chern-Simons formulation of 3D gravity, and its localization technique, 
for the holomorphic Lagrangian ${\cal{L}}$,  
\begin{align}
\label{ktlagrangian}
{\cal{L}}
&=\, \frac{i k}{4 \pi} 
\int_{\mc{M}} \text{Tr} \Big( \mc{A}d \mc{A} + \frac{2}{3} \mc{A}^3 
 - \Lbar{\lambda} \lambda + 2 D \sigma
\Big) \, \quad \nonumber \\
& \,\, \,\,+ t \int_{\mc{M}} \, 
 \text{Tr} \Big( 
\frac{1}{4} \mc{F}_{\mu \nu} \mc{F}^{\mu \nu}
+
\frac{1}{2} D_\mu \sigma D^\mu \sigma
+
\frac{1}{2} (D + \frac{\sigma}{l} )^2 \nonumber \\
&\,\,\, \,\,\, \,+
\frac{i}{4} \bar{\lambda} \gamma^\mu D_\mu \lambda
+
\frac{i}{4} \lambda \gamma^\mu D_\mu \bar{\lambda}
+
\frac{i}{2} \bar{\lambda} [\sigma , \lambda]
-
\frac{1}{4l} \bar{\lambda} \lambda
\Big) ,
\end{align}
and by summing over the geometries with Rademacher sum regularization \cite{rademacher:1939}. 
The need for Rademacher sum is due to the fact that, for gravity path-integral, 
one needs to sum over all the geometries consistent with localization locus, $\mc{F}_{\mu\nu} = 0$, but 
$\mc{F}_{\mu\nu} = 0$ is nothing but the Einstein's equation.  
The final expression for the holomorphic partition function, $Z_{hol}(q)$,  becomes 
\begin{align}
Z_{hol}(q)
=
R^{\left(- {k_{\text{eff}}}/{4}\right)} (q) -
R^{\left(- {k_{\text{eff}}}/{4} +1\right)} (q) \,,
\label{ptn}
\end{align}
where $q \equiv e^{2 \pi i \tau}$ with $\tau$ the complex moduli for the boundary torus, $k_{\text{eff}} = k+2$, $k = {\ell}/{4G_N}$ with $\ell$ as an AdS scale, and the function $R^{(m)} (q)$ is defined as 
\begin{align}
R^{(m)} (q)
&\equiv e^{2 \pi i m \tau} + \sum_{ \substack{ c > 0,   \\ (c,d)=1}}
\left( e^{2 \pi i m \frac{a \tau + b}{ c \tau + d}} -   r(a, b, c, d) 
\right) \nonumber \\
&=
q^m
+(\text{const})
+ \sum_{n=1}^\infty c(m,n)q^n
\,, 
\label{R}
\end{align}
where  $r(a, b, c, d) = e^{2 \pi i m \frac{a}{ c}}$ is $\tau$-independent 
quantity to regularize $c$, $d$ sum, and 
the coefficients $c(m,n)$ for the positive powers of $q$ are given in \cite{Iizuka:2015jma}. 

For later purpose, we express $c(m,n)$ in terms of modified Bessel function $I_1(z)$, 
\begin{align}
I_1(z) = \frac{z}{2} \sum_{\nu=0}^\infty \frac{ \left( \frac{z^2}{4} \right)^\nu }{\nu! \left( \nu+1 \right)!} \,,
\end{align}
as 
\begin{align}
\label{defofcmn}
&c(m,n)
=
2 \pi \sqrt{\frac{-m}{n}} \sum_{c=1}^\infty \frac{A_c (m,n)}{c} I_1 \Big( \frac{4 \pi \sqrt{-mn}}{c} \Big)
\,,
\end{align}
where $A_c(m,n)$ is so-called Kloosterman sum, 
\begin{align}
\label{Kloosterman}
A_c (m,n) =
\sum_{
\substack{
1 \leq d \leq c 
\\
(c,d)=1
}
}
e^{2 \pi i (m \frac{a}{c} + n \frac{d}{c})} \,. 
\end{align}
The summation in \eqref{Kloosterman} is for $d$, given the $c$, where $a $ 
is uniquely determined up to $(a, b) \approx (a + c, b + d)$ to satisfy $a d - b c = 1$ given $c$ and $d$ satisfying $(c,d)_{GCD} = 1$. 

The (const) term in \eqref{R} is undetermined, because it depends on the regularization scheme for the modular sum, namely how to choose $\tau$-independent quantity $r(a,b,c,d)$, and $r(a, b, c, d) = e^{2 \pi i m \frac{a}{ c}}$ for Rademacher sum is 
just one possible regularization scheme. 
In direct gravity calculation, there is apparently no principle to choose the right scheme. 
In this paper, we set the (const) term in \eqref{R} to be zero for convenience, though we admit that this is an open issue.  

One of the outstanding results in \cite{Iizuka:2015jma} is that 
for $k_{\text{eff}} =4$ in which the central charge $c_Q$ is expected to be 24, 
$Z_{hol}(q)$ becomes the $J$-function up to a constant shift which depends on the modular sum regularization scheme. 
Therefore, this quantum gravity partition function agrees with the 
extremal CFT partition function of Frenkel, Lepowsky, and Meurman \cite{FLM}, 
predicted by Witten in \cite{Witten:2007kt}.  
Note that in the case $k_{\text{eff}}=4$, $G_N \sim \ell$, therefore we are in full quantum gravity parameter regime.  

On the other hand, what happens when $G_N \ll \ell$? 
In this parameter regime, the semi-classical gravity description is guaranteed.  
The purpose of this short letter is, based on the result in \cite{Iizuka:2015jma}, to clarify the various physical aspects of the partition function in the semi-classical limit, where $G_N \ll \ell$.

\section{Boundary Fermion}
As is seen from \eqref{ptn} and \eqref{R}, the holomorphic partition function $Z_{hol}(q)$ has a negative pole in the large $k$ limit.  
Let us first comment on the physical origin of this minus sign: 
we claim that this is due to the fact that our partition function $Z_{hol}(q)$ contains additional fermion degrees of freedom, in addition to 
3D gravity. However, we can extrapolate 3D pure gravity partition function at least in the large $k$ limit, which  
we will explain now. 

First of all, the computation for \eqref{ptn} is based on the localization of $\mathcal{N}=2$ supersymmetric Chern-Simons theory  \cite{Iizuka:2015jma}, where 
we supersymmetrize the Chern-Simons formulation of the pure gravity by introducing auxiliary gauginos and scalars. 
The supersymmetrized action contains a ``mass term" for the gaugino, $\frac{k}{4 \pi}  \bar{\lambda}\lambda$.
Furthermore, for the localization we also added a supersymmetric exact term, {\it i.e.,} a super Yang-Mills term, which has a kinetic term for the gaugino, $\frac{t}{4} \bar{\lambda} \gamma^\mu D_\mu \lambda$. Note that we are forced to introduce the kinetic term even for the auxiliary field for localization.  
Furthermore in order to preserve some supersymmetries at the AdS boundary, which is equivalent to the boundary torus, 
we imposed the following boundary condition for the gauginos \cite{Sugishita:2013jca}
\begin{align}
\label{bdrycond}
\lambda\Big|_{\text{bdry}} = e^{- i \left( \varphi - t_E \right)} \gamma^3 \bar{\lambda}\Big|_{\text{bdry}},
\end{align}
where $e^{- i \left( \varphi - t_E \right)}$ is a phase factor depending on the torus coordinates, $t_E$ and $\varphi$ and $\gamma^3 = \left(
    \begin{array}{cc}
      1 & 0 \\ 
      0 & -1 
    \end{array}
  \right)$.  
Since the boundary condition \eqref{bdrycond} is incompatible with the gaugino ``mass term", 
this leads to the conclusion that the mass term vanishes at the boundary and therefore there is a boundary localized fermion.

If we look at the kinetic term and the mass term in the Lagrangian for the auxiliary fermions $\lambda$ and $\bar{\lambda}$, one can see that by using the doubling trick argument in \cite{Iizuka:2015jma}, they are like \begin{align}
\label{clasEOM}
{\cal L_{\lambda \,, \bar{\lambda}}} \sim t \, \psi_1 \Big(
\partial_{x} 
+ \frac{k}{t} \text{sign}(x) \Big) \psi_2,
\end{align}
where auxiliary fermions $\psi_1$ and $\psi_2$ are some 
components of $\lambda$ and $\bar{\lambda}$ (see \cite{Iizuka:2015jma} for detail) and  $x=0$ corresponds to the boundary. 
Then it is easy to see that this admits boundary localized fermion wave function as 
\begin{align}
\psi_2 (x, \varphi, t_E) = e^{- \frac{k}{t} |x|} \psi_2^{\text{boundary}} (\varphi, t_E) \,,
\label{bferm}
\end{align}
and that $\psi_2^{\text{boundary}}$ is sharply peaked at $x = 0$  
in the $t \to 0$ limit. 
Note that \eqref{clasEOM} is a classical Lagrangian for the auxiliary fermion, therefore the 
analysis of \eqref{bferm} from \eqref{clasEOM} and the conclusion that fermion is sharply localized at $x= 0$, 
are justified only in the classical limit, 
which is, in the large $k$ limit. 
This also implies that at finite $k$, 
there is no reason that the auxiliary fermion admits boundary localized mode. 
In fact, the analysis of \cite{Iizuka:2015jma} strongly suggests that there is no boundary localized fermion 
in the case of $k_{eff} = 4$, where central charge is expected to be 24. 
This is because then $Z_{hol}$ itself becomes the conjectured $J$-function by Witten \cite{Witten:2007kt}.

Following these, we claim that the boundary fermion 
contributes to the partition function as 
\begin{align}
Z_{\text{B-fermion}}(q)
=
\prod_{n=1}^\infty (1-q^n) \,.  
\label{ZBfermion}
\end{align}
Since this boundary localized fermion decouples at least from the bulk gravity in the large $k$ limit, 
and therefore it should give a common contribution to the partition function given by \eqref{ZBfermion}, independent of the bulk geometry. 
Then  
we claim that the quantity,  
\begin{align}
\frac{Z_{hol}(q)}{Z_{\text{B-fermion}}(q)} \equiv Z_{\text{gravity}}^{\text{large }k}(q)
\label{Zgrav}
\end{align}
represents physically ``bulk pure gravity'' partition function, at least in the semi-classical limit at large $k$.  
For later use, let us write down our holomorphic partition function \eqref{ptn} explicitly, 
\begin{align}
\label{Zhol}
Z_{hol}(q)
& \,= \,\,
\left(1-q \right) q^{- \frac{k_{\text{eff}}}{4} } 
+ \sum_{\Delta=1}^\infty
c_\Delta^{(k_{\text{eff}})}
q^\Delta \,, 
 \\
c_\Delta^{(k_{\text{eff}})}
&\, \equiv \,\,
c\left(- \frac{k_{\text{eff}}}{4} , \Delta \right) 
- c\left(- \frac{k_{\text{eff}}}{4} +1 , \Delta \right) \,.
\label{coef}
\end{align}
%
\section{Large $k$ pure gravity partition function}
Combining the equations \eqref{ZBfermion}, \eqref{Zgrav}, \eqref{Zhol} all together, we obtain 
\begin{align}
&Z_{\text{gravity}}^{\text{large }k}(q)
\nonumber \\
&\,=\,  q^{-\frac{k_{\text{eff}}}{4}} \prod_{n=2}^\infty \frac{1}{1-q^n}
+
\sum_{\Delta=1}^\infty
c_\Delta^{(k_{\text{eff}})}
q^\Delta
\prod_{n=1}^\infty \frac{1}{1-q^n} \,.
\label{gptn}
\end{align}
Note that the first terms in \eqref{gptn} is the usual Verma module made from the vacuum 
$| 0 \rangle$, satisfying
\begin{align}
L_{n>0}| 0 \rangle
=
L_{0}| 0 \rangle
=
L_{-1}| 0 \rangle
=0 \,,
\end{align}
and its descendants
$
\prod_{n \geq 2} L_{-n}^{N_n} | 0 \rangle \,. 
$
Therefore this is the Virasoro character for the vacuum ${Z}_{\text{vac}}(q)$,  
\begin{align}
q^{-\frac{k_{\text{eff}}}{4}} \prod_{n=2}^\infty \frac{1}{1-q^n} 
=
\text{Tr}_{  V_{|0 \rangle} }
q^{L_0 - \frac{k_{\text{eff}}}{4}} 
\equiv {Z}_{\text{vac}}(q) \,.
\end{align}
Similarly, the second term in \eqref{gptn} 
is the Verma module made from the primary state $| \Delta \rangle$, satisfying
\begin{align}
L_{n>0}| \Delta \rangle
=
0,
\quad
L_{0}| \Delta \rangle
=
(\Delta + \frac{k_{\text{eff}}}{4}) | \Delta \rangle \,,
\end{align}
and its descendants
$
\prod_{n \geq 1} 
L_{-n}^{N_n} | \Delta \rangle \,, 
$
with positive integer $\Delta$. Note that $L_{- 1}$ is included since $ | \Delta \rangle$ is not vacuum. 
Therefore this is the Virasoro character for generic primary ${Z}_{\text{primary}}(q)$, 
\begin{align}
q^\Delta
\prod_{n=1}^\infty \frac{1}{1-q^n} 
= 
\text{Tr}_{  V_{|\Delta \rangle} }
q^{L_0 - \frac{k_{\text{eff}}}{4}} 
\equiv {Z}_{\text{primary}}(q) \,. 
\end{align}
In the end we can write \eqref{gptn} as 
\begin{align}
Z_{\text{gravity}}^{\text{large }k}(q)
&\,=\,
{Z}_{\text{vac}}(q)
+
\sum_{\Delta=1}^\infty
c_\Delta^{(k_{\text{eff}})}
{Z}_{\text{primary}}^{\Delta} (q) \,.
\label{Zfamily}
\end{align}
This is our main point. 
The holomorphic partition function $Z_{hol}[q]$ given by \eqref{ptn} can be written as direct product of partition function from the boundary fermion \eqref{ZBfermion}  and one from bulk pure gravity \eqref{Zgrav} in the large $k$ limit. 
Furthermore, the partition function from pure gravity contribution \eqref{Zgrav} can be written as the sums over conformal families as \eqref{Zfamily}. 

This strongly suggests that there is a dual 2D CFT in the large $k$ limit 
and the coefficients $c_\Delta^{(k_{\text{eff}})}$ represent the number of  
primary operators labelled by $\Delta$ in the dual CFT. 
Clearly for that nice CFT interpretation possible, all $c_\Delta^{(k_{\text{eff}})}$ need to be positive integers. 
The fact that it is integer is easily confirmed step by step, following its definition  \eqref{coef}. 
We can also show that it is positive. However 
a few comments are worth making before we go into that. 

The fact that in the large $k$ limit, 
there is a huge gap of order $k_{eff}/4 + 1 \approx k/4 = \ell/16 G_N$ for conformal dimensions between vacuum and primary operators are 
exactly the half of the gap between AdS vacuum and BTZ black hole \cite{Banados:1992wn}. Half is because there is a same amount of contribution from the anti-holomorphic part.  
This gap is also the reason why Witten conjecture in \cite{Witten:2007kt} that the dual CFT, if exists, 
should be an extremal CFT, where BTZ black holes correspond to primary states. 
    
Another point is that $Z_{\text{B-fermion}}$ in \eqref{ZBfermion} is not modular invariant, so is 
$Z_{\text{gravity}}^{\text{large }k}(q)$ in \eqref{Zgrav}. 
One might wonder if modular non-invariant quantity is physical or not. 
Note that in the large $k$ limit, this modular invariance will be lost. 
This is because in the semi-classical limit, only the dominant saddle point survives and the rest saddle points vanish exponentially. 
However these vanishing saddle points are important ingredients for modular invariance property. 
Therefore we claim that only in the semi-classical limit, $k \to \infty$, quantity $Z_{\text{gravity}}^{\text{large }k}(q)$ \eqref{Zgrav} becomes physically meaningful since only in that limit, 
boundary fermion decouples and quantity $Z_{\text{B-fermion}}(q)$ becomes meaningful.  

\section{Positivity of $c_\Delta^{(k_{\text{eff}})}$}

We now show the positivity for $c_\Delta^{(k_{\text{eff}})}$. 
Since $c_\Delta^{(k_{\text{eff}})}$ in \eqref{Zfamily} is defined through \eqref{Zgrav} only in the large $k$ limit, 
we focus on $k_{eff} = k + 2  \to \infty$ limit in \eqref{coef}, 
which corresponds to the limit $m \to - \infty $ in $c(m,n)$ given by \eqref{defofcmn}. 
Since asymptotic behaviour of the modified Bessel function $I_\nu(z)$ is given by the exponential 
growth 
\begin{align}
\lim_{z \to \infty} I_\nu (z) \sim \frac{ e^{z} }{ \sqrt{ 2 \pi z} } \left( 1 - \frac{4 \nu^2 - 1^2}{8 z} + {\cal{O}}(z^{-2})\right) \,, 
\end{align}
in the summation over $c$ expression for $c(m,n)$ in \eqref{defofcmn}, $c=1$ contribution dominates and 
the contributions $c > 1$ are suppressed exponentially.  
In $c=1$ case,  for any integer values $m$ and $n = \Delta$,  
\begin{align}
A_{c=1}(m, \Delta) = 1
\end{align} 
is easily seen from  \eqref{Kloosterman}. 
Note that $k_{eff}/4$ needs to be positive integer \cite{Iizuka:2015jma}, and this 
gives $m$ to be negative integer. 
Therefore, the dominant contribution in the summation expression for $c(m,n)$ in \eqref{defofcmn} is 
given by 
\begin{align}
& \lim_{m \to - \infty} c(m,n) 
= \frac{\,\,(-m)^{1/4} }{2^{1/2}  n^{3/4}} e^{4 \pi \sqrt{-m n}} 
\Bigl( 1  - \frac{3}{32 \pi \sqrt{- m n}} 
\nonumber \\
& \qquad \qquad \qquad \quad  + {\cal{O}}\left({-m n}\right)^{-1}  \Bigr) 
 + {\cal{O}}(e^{2 \pi \sqrt{-m n}}  ) \,.
\end{align}
Then, from \eqref{coef} it is clear that 
\begin{align}
 \lim_{k \to \infty} 
 c_\Delta^{(k_{eff})} =& \, \, \frac{2 \pi }{       ( {k_{eff}} )^{1/4} \Delta^{1/4}  } 
e^{2 \pi \sqrt{ k_{eff} \Delta}}  
 \left( 1 + {\cal{O}}(  k^{-{1}/{2}}_{eff} ) \right)  \, \nonumber \\
  >& \, \, 0 \,,  
\label{cposi}
\end{align}
and this leads to, in the leading order,  
\begin{align}
\label{Cardyformula}
\lim_{k \to \infty} \log c_{\Delta}^{(k_{eff})} = 2 \pi \sqrt{k_{eff} \Delta} = 2 \pi \sqrt{ \frac{c_{Q} \Delta }{6}} \,,
\end{align}
where $c_{Q}$ is for central charge and we have used the relationship $c_Q = 6 k_{eff}$.  
The result \eqref{Cardyformula} perfectly matches with the boundary CFT's Cardy formula in the large central charge $c_{Q}$  limit. 

Since it is defined only in the large $k$ limit through the definition \eqref{Zgrav}, 
the expression \eqref{Zfamily} is meaningful only in the large $k$ limit. However 
one can also easily check that $c_\Delta^{(k_{\text{eff}})}$ is positive and also integer 
by using the definition \eqref{coef} in a straightforward way numerically for finite $k$ case, where  
our coefficients with analytic method 
agree with numerics. 
But we take other approach here: 
we write $R^{(m)}(q)$ in terms of polynomial 
of $J$-function as \cite{Witten:2007kt} 
in order to know the $c(m,n)$ order by order, where $J$ is given in terms of Klein's $j$-invariant as $J = j - 744$.  
This is possible since both $R^{(m)}(q)$ and positive powers of $J$-function are modular invariant and have no pole other than $q=0$.  
For example,  
\begin{align}
& R^{(-1)}(q) = J(q) 
\notag \\
&= \frac{1}{q} + \underbrace{196884}_{c(-1,1)} q + \underbrace{21493760}_{c(-1,2)} q^2 + \dots,
\notag \\
& R^{(-2)}(q) = J(q)^2 - 393768
\notag \\
&=
\frac{1}{q^2} + \underbrace{42987520}_{c(-2,1)} q + \underbrace{40491909396}_{c(-2,2)} q^2 + \dots,
\notag \\
& R^{(-3)}(q) = J(q)^3 - 590652 J(q) - 64481280
\notag \\
&=
\frac{1}{q^3} + \underbrace{2592899910}_{c(-3,1)} q + \underbrace{12756069900288}_{c(-3,2)} q^2 + \dots,
\end{align}
and so on. 
As one can see, the coefficient $c(m,n)$ growing very fast with respect to $-m$ with fixed $n$. Therefore, 
this yields positivity for $c_{\Delta }^{(k_{eff})}$.  
One can observe, 
for example, 
\begin{align}
&c_{\Delta = 1}^{(4)} = 196884,
\quad
c_{\Delta = 2}^{(4)}= 21493760 , \nonumber 
\\
&c_{\Delta = 1}^{(8)}= 42790636,
\quad
c_{\Delta=2}^{(8)} = 40470415636, \nonumber 
\\
&c_{\Delta = 1}^{(12)}= 2549912390,
\quad
c_{\Delta =2}^{(12)} = 12715577990892,
\end{align}
from the above numerics.
We observed that positivity of $c_\Delta^{(k_{\text{eff}})} > 0$ holds 
up to $k_{\text{eff}}/4 = 30$, $\Delta = 50$.  
In these ways, even in the finite $k$ region, 
we can 
conduct step by step check for the positivity and integer nature, 
and as we have already shown in \eqref{cposi}, in the large $k$ parameter region, 
 it is positive. 

\section{Summary and discussion}
In this short paper, we focused and analyzed the various physical aspects of our previous results \cite{Iizuka:2015jma} in the large $k$ limit. 
Large $k$ corresponds to, through the relation $\ell/G_N \propto k$, the semi-classical limit of quantum gravity. 
We obtain an plausible representation
of our 
partition function \eqref{gptn} just by dividing the original our holomorphic partition function by the contribution coming from the boundary-localized fermion. 
This suggests that our full quantum gravity partition function \eqref{ptn} contains contributions from both bulk gravity and the boundary fermion, and in the dual CFT, they couple in the finite $k$, but only in the large $k$ limit, they decouple. 
The gravity partition function, obtained by dividing the boundary fermion contribution as \eqref{Zgrav}, is written as summation over 
Virasoro characters for the vacuum and primaries as \eqref{Zfamily}. We have shown that the coefficients $c_\Delta^{(k_{eff})}$ in \eqref{Zfamily}
are positive definite integers and satisfy the Cardy formula in the large $k$ limit. These facts give a consistency check to interpret \eqref{Zfamily} as a
dual 2D CFT partition function.

In this paper, we claim that in nonperturbative formulation of 3D quantum ``pure'' gravity, 
involving additional fermion degrees of freedom is unavoidable. 
One might wonder why we cannot obtain non-perturbative partition function for just pure gravity, without any 
additional degrees of freedom. Let us discuss this point in detail now. 
Our claim is that, in order to conduct the metric path integral exactly at the nonperturbative level, 
it is better to use the localization technique and for that,  
``$t$ regularization'' \footnote{Here we mean `$t$ regularization' as deforming the original theory by addition of 
$t {\cal{L}}_{SYM}$ term as \eqref{ktlagrangian}.}. 
Otherwise, one has to rely on the perturbative 
analysis, unless one can solve it exactly.  

About exact solvability, 
since Chern-Simons theory is topological, it could be that one can solve it exactly, without relying 
on perturbation. 
In \cite{Elitzur:1989nr}, purely bosonic Chern-Simons theory is discussed in detail. 
Therefore by using the results of \cite{Elitzur:1989nr}, one might be able to obtain 
$Z_{(c,d)}$ in our previous paper \cite{Iizuka:2015jma} without relying 
localization. 
However, even if one could, 
it gives at most the non-perturbative results for the Chern-Simons theory with fixed topology only. It 
does not give the non-perturbative results for quantum gravity, {\it i.e.,} the justification for the emergence of modular/Rademacher sum. 

About perturbative analysis, it is pointed out in several literatures \cite{Maloney:2007ud,Giombi:2008vd} that 
gravity path integral is ``one-loop exact''. 
However there is no direct calculation to confirm this solely in gravity side. Many literatures  
assume some properties motivated by dual CFT. 
Furthermore, even if the one-loop exactness is true, it is at most perturbative level.  
On the other hand, by $t$ regularization for localization, 
we succeeded in conducting the exact path integral for the metric nonperturbatively, and 
as a result, we obtain full partition function $Z_{hol}(q)$. Note 
that in our calculation, 
we do not assume the existence of dual CFT, nor one-loop exact, nor Virasoro algebra.   
In this paper, we propose how to extrapolate pure gravity contribution at least in the large $k$ semi-classical limit, and 
as a result of our calculation, we obtain the Virasoro characters. 

Furthermore our localization calculation gives very naturally 
the reason for the emergence of modular/Rademacher sum, 
very important non-perturbative effects:  
this is because in localization calculation, only the localization locus 
$\mc{F}_{\mu\nu} = 0$ contributes in the path-integral but we have to sum over 
all of the field configurations satisfying $\mc{F}_{\mu\nu} = 0$. Since $\mc{F}_{\mu\nu} = 0$ 
is nothing but the 
Einstein's equation written in terms of the holomorphic gauge field $\mc{A}_\mu$, 
summing over all the field configurations satisfying $\mc{F}_{\mu\nu} = 0$ exactly corresponds to 
summing over all of the {\it complex}
solutions of the Einstein's equation.  Complex is due to the holomorphic property. 
This corresponds to  summation over `cosets of $SL(2, Z)$', 
{\it i.e.,} the summation over $c$ and $d$ for the Rademacher sum,  
because $c$ and $d$ characterise how to embed all of the complex solutions of the Einstein's equation into solid torus, and therefore, characterise all the complex valued saddle points.  
Thus, in our localization calculation, we do {\it not} have to impose summing over `cosets of $SL(2, Z)$' {\it by hand}, 
rather it arises naturally as localization locus from the exact path-integral \footnote{Precisely speaking, for this to work, we have to modify the meaning of Chern-Simons theory path integral to more appropriate one for gravity: we have to sum over all of the solutions of $\mc{F}_{\mu\nu} = 0$ whose boundary are related by cosets of $SL(2, Z)$. Summing over different boundary conditions related by cosets of $SL(2, Z)$ is not what we do for the Chern-Simons theory, but we need that since they give physically the same boundary condition in gravity.}. 
Furthermore, this gives the explanation why geometries like singular ones do not contribute to the path integral.  This is simply because they are not the localization locus, satisfying $\mc{F}_{\mu\nu} = 0$.

Note that even though 
the partition function is written as the direct product of holomorphic and anti-holomorphic one, and therefore 
there are complex valued saddle points parametrised by $c$ and $d$, 
in the semi-classical limit the dominant saddle point gives 
the real valued saddle point, see \S 4.2 of \cite{Maloney:2007ud}. 
Similarly Hawking-Page transition also occurs even for such holomorphic partition function.

Finally it is very interesting to generalize 
the results we obtained in this paper to higher spin gravity. In fact, it can be shown that 
our good expression for the partition function, where it is expressed as summation over vacuum and primary characters, 
holds even in higher spin gravity \cite{WIP}. 
It would be very interesting to understand quantum gravity in AdS$_3$ in more great detail through these generalization. 
We will 
come back to these issues in near future.

\vspace{.5cm}
\acknowledgments
AT would also like to thank Osaka University for its hospitality where part of this work was done. 
ST would also like to thank M. Shigemori for discussion. 
The work of NI was supported in part by 
JSPS KAKENHI Grant Number 25800143. 
The work of AT was supported in part by the RIKEN iTHES Project.





\end{document}